**Caught in the Crossfire: Fears of Chinese-American Scientists**

Yu Xie[1]*

Xihong Lin[2]

Ju Li[3]

Qian He[1]

Junming Huang[1]

Word Count: 3,387.

_______________________________

1. Center on Contemporary China, Princeton University, Princeton, NJ 08544. 2.
2. Department of Biostatistics and Department of Statistics, Harvard University, 655 Huntington Avenue, Boston, MA 02115.
3. Department of Nuclear Science & Engineering and Department of Materials Science & Engineering, Massachusetts Institute of Technology, 77 Massachusetts Avenue, Cambridge, MA 02139
* Corresponding author: yuxie@princeton.edu.

## Caught in the Crossfire: Fears of Chinese-American Scientists

### Abstract

The US leadership in science and technology has greatly benefitted from immigrants from other countries, most notably from China in the recent decades. However, feeling the pressure of potential federal investigation since the 2018 launch of the China Initiative under the Trump administration, Chinese-origin scientists in the US now face higher incentives to leave the US and lower incentives to apply for federal grants. Analyzing data pertaining to institutional affiliations of more than 2.3 million scientific papers, we find a steady increase in the return migration of Chinese-origin scientists from the US back to China. We also conducted a survey of Chinese-origin scientists employed by US universities in tenure or tenure-track positions (n=1300), with results revealing general feelings of fear and anxiety that lead them to consider leaving the US and/or stop applying for federal grants.

### Significance Statement

Our study reveals the widespread fear among Chinese-origin scientists in the US arising from conducting routine research and academic activities. If this fear is not alleviated, there are significant risks of an underutilization of scientific talent as well as losing scientific talent to China and other countries. Addressing the fear of Chinese-origin scientists and making the American academic environment more welcoming and attractive to them will help retain and attract scientific talent and strengthen the US leadership in science and technology in the long run.

**Caught in the Crossfire: Fears of Chinese-American Scientists**

A 2007 report, *Rising above the Gathering Storm* (*1*)*,* shocked the scientific community with an alarming message that American science may be in decline and soon lose its long-held leadership in the world. Evidence cited in support of this claim included inadequate US investments in science education at all levels and in scientific research, in an era when competing countries, China in particular, had been increasing science-related investments and narrowing gaps with the US. This report received a great deal of attention from policymakers, spawning over two dozen bills in Congress within a year of its release.

Addressing this science policy question, sociologists Xie and Killewald published a book in 2011, *Is American Science in Decline?* (*2*)*.* After examining a variety of indicators on science, Xie and Killewald dismissed the alarmist view of the 2007 report and concluded that American science had fared reasonably well. One of the main reasons for their relatively optimistic conclusion was America's benefit from immigration: even if the US does not train an adequate number of scientists and engineers that it needs for its modern economy, it is able to attract the best and the brightest scientists and engineers from around the world. For example, China has been the most important foreign supplier of US-based scientists for several decades.

**Chinese Scientists in the US**

Out of about 34,000 Ph.D. recipients in science/engineering (S/E) fields awarded by US institutions in 2020, 46% (approx. 15,000) held temporary visas, a lower-bound estimation of "foreign students." Among these 15,000 recipients with temporary visas, the largest portion came from China, at 37%. In other words, 17% of all 2020 US doctoral degrees in S/E went to foreign students from China (see Supplementary Materials 1, Table S1). Most foreign-born noncitizen recipients of US S/E doctorates remain in the US for subsequent employment. For those from China, about 87% have stayed in the US, constituting a significant part of the American S/E labor force (Supplementary Materials 1). Along with native-born Chinese Americans, Chinese immigrants have become a large and visible demographic group in American science and technology (*3*). Today, it is hard to open an issue of any major scientific journal and not to find a Chinese name among its contributing authors. However, both the future supply and retention of current scientists and engineers from China have been impacted by the chilling effect of the "China Initiative" launched by the US federal government in 2018.

**The China Initiative**

In 2018, the Department of Justice under the Trump administration launched the China Initiative with the objective of stopping "Chinese economic espionage" (*4*). In reality, the Initiative mostly targeted US-based academic scientists of Chinese origin for "research integrity" issues, the most prominent being failure to disclose relationships with Chinese institutions on federal grant applications, particularly those to the National Institutes of Health (*5*). The Initiative was heavily criticized for its ethnic profiling tactics by both the scientific community and civil rights advocates, leading to an ending of its official name in early 2022, but not its substantive operations (*6*). So far, the China Initiative has openly investigated about 150 academic scientists and prosecuted two dozen of them with criminal charges (*5, 6*).

One high-profile case was against Gang Chen, a former head of the Department of Mechanical Engineering at MIT and a member of the US National Academy of Engineering. After his arrest on January 14, 2021, his lab was closed, and his research group dispersed. A year later, all charges were dropped (*7*). The chilling effect of the Gang Chen case was significant and consequential; it resulted in greater community awareness among Chinese-American scientists and heralded nationwide discussions in the community as to how to protect oneself. For example, a new non-profit organization, the Asian American Scholar Forum (AASF), was established in response to Gang Chen's case to promote academic belonging, openness, freedom, and equality for all. Since 2021, three surveys of Chinese-American scientists have been conducted to understand their concerns and feelings in this new climate (Supplementary Materials 4, 8).

**The Reverse Brain Drain of Chinese Scientists in the US**

The China Initiative caused panic and an exodus of senior academic researchers of Chinese descent in the US. When Song-Chun Zhu, an accomplished computer scientist and the director of the Center for Vision, Cognition, Learning and Autonomy at UCLA, announced his intention to return to China in 2019, an article was widely circulated on Chinese social media, publicly thanking Donald Trump and his China Initiative for sending top Chinese-American scientists like Zhu back to China (*8*). Zhu currently serves as the dean of the Institute for Artificial Intelligence at Peking University.

While mainland China's contribution to the world's science and technology was minor only three decades ago, it is now a major contributor of science and technology (*9*). In terms of the total number of science and technology publications in scientific journals, China has now surpassed the US as the world leader (*9*). In terms of patent applications by residents, China outperforms the US by a factor of five (Supplementary Materials 2). Four explanations accounting for China's recent success in science and technology development are (1) a large population and human capital base, (2) a labor market rewarding academic meritocracy, (3) a centralized government willing to invest in science, and (4) the return migration of foreign-trained scientists and engineers of Chinese origin to China *(10)*. Chinese-origin scientists living and working overseas have been lured to return to China by a combination of factors: large and fast-growing investments in science, high social prestige and attractive financial rewards tied to positions in Chinese institutions, and capable research collaborators and assistants. In this study we ask whether and to what extent—net of these "pull" factors—the China Initiative contributed to pushing Chinese-origin scientists to return to China.

We conducted an analysis to estimate trends in return migration of Chinese-origin scientists to China using bibliometric data. The methodology is described in Supplementary Materials 3. The trends, respectively, for life science, mathematics and physical science, and engineering and computer science, are presented in Figure 1, separately for junior scholars (Figure 1a) and experienced scholars (Figure 1b). We define experienced scholars as those with 25 or more publications (see Supplementary Materials 3). The Y-axis represents the ratio of the number of returning scientists each year relative to the baseline in 2005–2010 by corresponding fields. It is apparent that the number of returning scientists had been increasing steadily before the China Initiative, and that this was true for both junior scholars and experienced scholars.

By 2018, the factor ranged between 4 and 5 for junior scholars and 3 and 4 for experienced scholars, across each of the fields. After 2018, when the China Initiative was first implemented, the trend picked up speed, reaching the 5–6 range in 2021, except for life scientists. While the return rate slowed for junior life scientists, it increased for experienced life scientists after 2019. This finding is consistent with the reported sharp fall in dual affiliations and collaborations between the US and China by 2021 (*11*).

**Fears of Chinese-American Scientists**

Relative to the size of the total Chinese-American scientist/engineer population, the number who have returned to China is very small. The vast majority prefer to stay and continue their work in the US. However, they now fear that their work and lives in the US may be jeopardized by the China Initiative.

Between December 2021 and March 2022, we conducted an online survey of US-based scientists of Chinese origin on behalf of the AASF. We obtained responses from 1,304 Chinese-American researchers currently employed by US universities. They are well represented in terms of geography, institution type (private versus public), gender, field of study, and seniority (Supplementary Materials 4). By survey standards, the AASF survey is a "convenience" sample. It is not a probability-based sample because there is no national sampling frame from which we could draw such a sample. In Supplementary Materials 9, we compare the representativeness of the sample with data from the American Community Survey (ACS).

A methodological caveat is in order. There are two sources of potential bias with our survey data (discussed in more detail in Supplementary Materials 4): "sample selection bias" and "social desirability bias," both in the direction of exaggeration of the negative impact of the China Initiative. Therefore, caution is needed when we interpret the results. However, the high degree of consistency of our survey results with those from two other similar surveys (Supplementary Materials 8) lends credence to the results we report below.

In Figure 2, we present our main findings with eight indicators: five "psychological indicators" and three "intention indicators." Our results are largely consistent with the findings from two earlier similar surveys (*12*). In Supplementary Materials 8, we compare both the design and the findings across the three surveys. All five psychological indicators reveal a strong sense of uneasiness and fear: 35% of respondents feel unwelcome in the US, and 72% do not feel safe as an academic researcher; 42% are fearful of conducting research; 65% are worried about collaborations with China; and a remarkable 86% perceive that it is harder to recruit top international students now compared to five years ago. The intention indicators address the potential impact of these psychological concerns on behavioral intent: 45% of respondents who have obtained federal grants say that they now wish to avoid applying for federal grants; and a shocking 61% have thought about leaving the US (for either Asian or non-Asian countries). Among those who intend to continue applying for federal grants, 95% indicate they rely on grants to conduct research, especially life scientists. Despite an overall fearful sentiment, an overwhelming majority (89%) of our respondents indicated their desire to contribute to the US leadership in science and technology.

Regression analyses predicting the first two behavioral intentions with demographic and professional characteristics, presented in Supplementary Materials 5 (Models 1A and 1B), reveal that faculty members in engineering and computer science, those of senior ranks, and those from public institutions are much more likely to consider avoiding federal grant applications. Our results also show that junior faculty and those who have been funded by federal grants are much more likely to consider relocating abroad. This is particularly worrisome because junior researchers and federal grant awardees are important to the global competitiveness of the US in cutting-edge science and technology.

As reported in Supplementary Materials 5 (Models 2A and 2B), we also find that indicators of fear (shown in Figure 2) strongly predict the first two intention measures—avoiding federal grant applications and considering relocating abroad, after adjusting for demographic, professional, and geographical covariates. Variables capturing perceptions of professional belonging and university leadership are not significantly predictive of those two intentions. After accounting for psychological indicators, engineering and computer science faculty are not statistically different from other respondents in avoiding federal grant applications, suggesting that fear of conducting research explains the observed difference. After accounting for these fear effects, junior faculty and federal grant awardees remain much more likely to consider leaving the US.

These survey data on Chinese-American scientists should be interpreted in the broader US context. Not only were Chinese-Americans subject to racial discrimination in America's past, anti-Asian and anti-Chinese sentiments in the US have increasingly prevailed since the COVID-19 pandemic began (*13, 14*). The high percentage of those considering leaving the US is partly attributable to a Chinese-hostile societal environment in the US nowadays. Our data show that 83% of the respondents had experienced insults in a non-professional setting in the past year, and experiencing insults of this kind significantly heightened individuals' intention of leaving the US. However, this large societal effect of insult experiences does not explain away the net effects of "fear" and "feeling unwelcome" resulting from the China Initiative on the intention of leaving the US.

We further explore the reasons behind our respondents' fears. Supplementary Materials 6 displays the detailed results. Our analysis suggests that engineering and computing science faculty, life science faculty, federal grant awardees, senior faculty, and males are relatively more likely to feel fearful of conducting research in the US.  Of the five possible reasons for "not feeling safe as an academic researcher in the US," most survey respondents pointed to fears of "US government investigations into Chinese-origin researchers" (67%) and "Anti-Asian hate and violence in the US" (65%). Meanwhile, relatively smaller percentages of respondents expressed other fears, such as that "US government officials often attack the Chinese government or Chinese policies" (38%), "My family, friends, or collaborators might be targeted by the US or Chinese government in retaliation for something I say or do" (37%), and "Others might report what I say or do to the US or Chinese government" (31%).

Our survey uncovers many Chinese-American scientists' intention to avoid applying for federal grants out of fear of federal government prosecution under the China Initiative. In our data, of the 445 respondents who intended to avoid applying for federal grants, 84% indicated that this was "Because I am afraid that I would have legal liability if I made mistakes in forms and disclosures," while 66% reported that this was "Because I worry that my collaborations with Chinese researchers or institutions would place me under suspicion."

Our survey instrument allowed our respondents to make open-ended comments at the end of the survey, yielding hundreds of comments. One respondent, self-identified as a US citizen and a former recipient of the National Science Foundation CAREER Award, told us that he quit his academic position exactly because of what he perceived as an "anti-Chinese atmosphere." He then wrote:

If it were not because the COVID pandemic cuts off international traveling and I am a U.S. citizen, my family would have left the U.S. permanently without any intent to come back in the future. What I have experienced at my former institution was not only disgusting, but a system[ic] corruption that I believe [is] illegal. I had never thought of somewhere in this county to be dark and corrupted like this. If I had, I would not have become a naturalized U.S. citizen, which I regret now. What I ha[ve] experienced not [only] ruined my academic career, but also destroyed my American dream.

**Conclusion**

Immigrant scientists and engineers from China have been an integral part of the US research enterprise for decades. In the past, there have been complaints that while they contributed a large share of the hard work, on the whole they failed to achieve leadership positions or commensurate recognition, reaching a "bamboo ceiling" (*15*, *16*). Under the China Initiative, a majority of Chinese-origin American scientists now feel the chilling effect of potential federal investigations and prosecution and have a new reason to be pessimistic about their careers in the US. Indeed, although an overwhelming majority would like to contribute to the US leadership in science and technology, many feel unwelcome and fearful of conducting research in the US. For some Chinese-American scientists, this fear leads to their consideration of avoiding federal grant applications, especially among engineering and computer science faculty, and of leaving the US, especially among junior faculty and federal grant awardees. There are indications that applications for National Science Foundation grants declined significantly between 2011 and 2020 (*17*). While the decline was 17% overall, it was much higher, at 28%, for Asian American scientists.

Modern science has been making tremendous progress since its inception in the seventeenth century because it has been open, benefitting the entirety of humanity. The world center of science has shifted several times in the past, from Renaissance Italy to England in the seventeenth century, to France in the eighteenth century, and to Germany in the nineteenth century, before crossing the Atlantic in the early twentieth to the US (*2*). Still, scientists everywhere have belonged to a single worldwide community, as they share new knowledge with one another through publications in the public domain. What attracts scientists the most is not material comfort but academic freedom and opportunities to pursue one's ideas; for a long time, the US has been providing a working environment that is more conducive to these values than that of any other country (*2*). This is and should remain a distinctive advantage of the US.

In this article, we have shown unintended consequences of the China Initiative that are harmful to American science: (1) discouraging new Ph.D. recipients of Chinese origin from working in the US, (2) encouraging world-class Chinese-American scientists to leave the US, especially junior researchers and federal grant awardees; and (3) discouraging experienced Chinese-American scientists from securing federal sponsorship, especially among engineering and computer science faculty. Addressing the fears of scientists of Chinese origin and making the academic environment welcoming and attractive for all will help retain and attract scientific talent and strengthen the US leadership in science and technology in the long run.

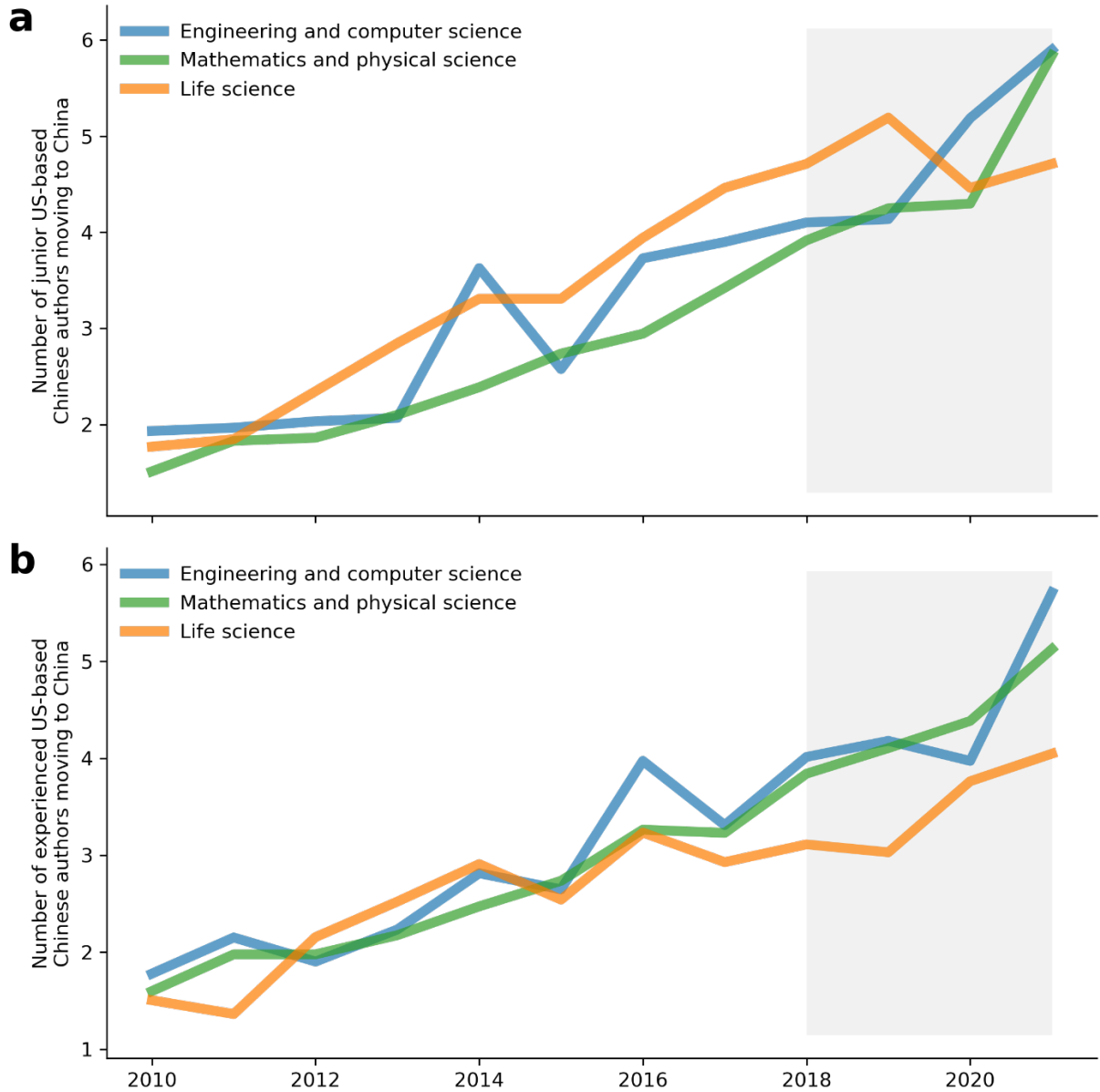

**Figure 1: Normalized number of (a) junior and (b) experienced Chinese scientists leaving the US each year for China from 2010 to 2021.** Note: Chinese scientists are counted as "leaving" if they published their first paper with an affiliation in the US and later published with a China affiliation but without an affiliation in the US. Yearly numbers are normalized by the average number of leaving Chinese scientists in 2005–2010 to ensure the reported numbers are comparable across disciplines. The shaded portion highlights the notable increase after 2018.

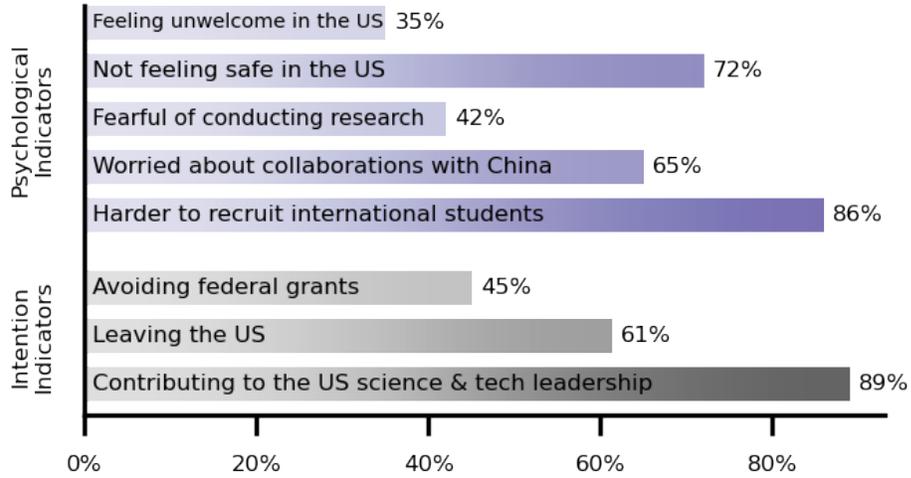

**Figure 2:    Chinese-origin scholars' perceptions and intentions.**  Note: Only past and current grant awardees were asked the question of whether they were considering "avoiding applying for federal grants."

Supplementary Materials for

**Caught in the Crossfire: Fears of Chinese-American Scientists**

Yu Xie[1*]

Xihong Lin[2]

Ju Li[3]

Qian He[1]

Junming Huang[1]

_______________________________

1. Center on Contemporary China, Princeton University, Princeton, NJ 08544.
2. Department of Biostatistics and Department of Statistics, Harvard University, 655 Huntington Avenue, Boston, MA 02115.
3. Department of Nuclear Science & Engineering, Massachusetts Institute of Technology, 77 Massachusetts Avenue, Cambridge, MA 02139
* Corresponding author: yuxie@princeton.edu.

**Supplementary Materials 1: Ph.D. Students from China**

We calculated the number of science and engineering (S/E) Ph.D. recipients and those of them holding temporary visas in the US in 2020 from data reported by the National Science Foundation (NSF) Survey of Earned Doctorates (*1*).  We aggregated the data in Table 17 ("Doctorate recipients, by broad field of study and citizenship status: Selected years, 1975–2020") in the Survey of Earned Doctorates data tables across four major fields: life sciences, physical sciences and earth sciences, mathematics and computer sciences, and engineering. We then obtained the number of S/E Ph.D. recipients from China in Table 26 ("Top 10 countries of origin of temporary visa holders earning US doctorates, by country of citizenship and field of study: 2010–20"). The numerical results are given in Table S1.

**Table S1: Number of S/E Ph.D. in 2020 by immigration status and Chinese origin**

|  | Numbers |
| --- | --- |
| Total | 33,676 |
|  |  |
| US citizen or permanent resident | 18,338 |
| Temporary visa holder from all countries | 15,338 |
| From China | 5,730 |

The NSF also reports "stay rates," percentages of US doctorate recipients holding temporary visas who intend to stay in the US by countries of origin (*2*).  For all temporary visa holders, the average stay rate in 2005–2015 was 73.7%. For those from China, the average stay rate was 87.2%.

**Supplementary Materials 2: Trends in Patent Applications, China versus the US**

We measure the yearly number of patent applications in China and the US as an indicator of

technology development in the two countries. The data are published by the World Bank and

collected by CEIC Data (*3*), for the following series: US Patent Applications: Residents (ID:

265144402), US Patent Applications: Non-Residents (ID: 265129602), Chinese (CN) Patent

Applications: Residents (ID: 265132202), and Chinese (CN) Patent Applications: Non-Residents

(ID: 265123902). The US data began in 1980, and the Chinese data began in 1985.

Table S2: Yearly number of patent applications in China and United States, by resident status.

|  | CN Patent Applications: Non-Residents | CN Patent Applications: Residents | US Patent Applications: Residents | US Patent Applications: Non-Residents |
|---|---|---|---|---|
| 1980 |  |  | 62,098 | 42,231 |
| 1981 |  |  | 62,404 | 44,009 |
| 1982 |  |  | 63,316 | 46,309 |
| 1983 |  |  | 59,391 | 44,312 |
| 1984 |  |  | 61,841 | 49,443 |
| 1985 | 4,493 | 4,065 | 63,673 | 51,562 |
| 1986 | 4,515 | 3,494 | 65,195 | 55,721 |
| 1987 | 4,084 | 3,975 | 68,315 | 63,522 |
| 1988 | 4,872 | 4,780 | 75,192 | 68,644 |
| 1989 | 4,910 | 4,749 | 82,370 | 76,337 |
| 1990 | 4,305 | 5,832 | 90,643 | 80,520 |
| 1991 | 4,051 | 7,372 | 87,955 | 84,160 |
| 1992 | 4,387 | 10,022 | 92,425 | 90,922 |
| 1993 | 7,534 | 12,084 | 99,955 | 84,241 |
| 1994 | 7,876 | 11,191 | 107,233 | 95,522 |
| 1995 | 8,688 | 10,011 | 123,962 | 104,180 |
| 1996 | 11,114 | 11,628 | 106,892 | 105,054 |
| 1997 | 12,102 | 12,672 | 119,214 | 101,282 |
| 1998 | 33,645 | 13,751 | 134,733 | 102,246 |
| 1999 | 34,418 | 15,626 | 149,251 | 116,512 |
| 2000 | 26,560 | 25,346 | 164,795 | 131,100 |
| 2001 | 33,412 | 30,038 | 177,513 | 148,958 |

| 2002 | 40,426 | 39,806 | 184,245 | 150,200 |
|------|--------|--------|---------|---------|
| 2003 | 48,548 | 56,769 | 188,941 | 153,500 |
| 2004 | 64,598 | 65,786 | 189,536 | 167,407 |
| 2005 | 79,842 | 93,485 | 207,867 | 182,866 |
| 2006 | 88,183 | 122,318 | 221,784 | 204,182 |
| 2007 | 92,101 | 153,060 | 241,347 | 214,807 |
| 2008 | 95,259 | 194,579 | 231,588 | 224,733 |
| 2009 | 85,508 | 229,096 | 224,912 | 231,194 |
| 2010 | 98,111 | 293,066 | 241,977 | 248,249 |
| 2011 | 110,583 | 415,829 | 247,750 | 255,832 |
| 2012 | 117,464 | 535,313 | 268,782 | 274,033 |
| 2013 | 120,200 | 704,936 | 287,831 | 283,781 |
| 2014 | 127,042 | 801,135 | 285,096 | 293,706 |
| 2015 | 133,612 | 968,252 | 288,335 | 301,075 |
| 2016 | 133,522 | 1,204,981 | 295,327 | 310,244 |
| 2017 | 135,885 | 1,245,709 | 293,904 | 313,052 |
| 2018 | 148,187 | 1,393,815 | 285,095 | 312,046 |
| 2019 | 157,093 | 1,243,568 | 285,113 | 336,340 |
| 2020 | 152,342 | 1,344,817 | 269,586 | 327,586 |

**Supplementary Materials 3: Trends in Migration of Chinese-American Scientists from the US to China**

We estimate the trends in the migration of US-based Chinese scientists to China by drawing on the large-scale academic bibliometrics database Microsoft Academic Graph (*4*), which indexed 208,440,142 scientists from 27,077 institutions authoring 2,316,278,852 scientific publications dated until December 2021.

We identified Chinese scientists by their surnames. We first collected 832 common Chinese surnames from Wikipedia (https://en.wikipedia.org/wiki/List_of_common_Chinese_surnames), including those in Chinese characters and romanized names, in Hanyu Pinyin (the system of Chinese romanization mostly used by mainland Chinese scientists) and Wade–Giles (the system mostly used by Cantonese-speaking and Taiwanese scientists). This methodology results in the non-counting of Chinese scientists who have changed their surnames (usually females after marriage), leading to an undercount.

We searched for those surnames in the authors' full names recorded in Microsoft Academic Graph, and identified a total of 28,140,577 Chinese scientists. To retain a high degree of reliability in individual identification, we removed scientists with a gap of more than 5 years between consecutive publications, which we believed were false results in which Microsoft Academic Graph's name disambiguation algorithm incorrectly merged multiple individuals. We ended up with 27,595,008 Chinese scientists.

Microsoft Academic Graph records every paper with one or more field labels from a total of 716,883 possible fields, such as "message passing" or "quantum process." Along with those labels comes a tree-like structure grouping small fields into 19 first-level fields and 292

second-level fields. We mapped all those first- and second-level fields to 4 major disciplines:

mathematics and physical science (including statistics), life science, engineering and computer

science, and social sciences and others, following the classification in Xie and Shauman's book

*Women in Science* (*5*).

Table S3: Grouping Microsoft Academic Graph fields into 4 major disciplines.

| Major disciplines | Microsoft Academic Graph first-level field | Microsoft Academic Graph second-level field |
|---|---|---|
| Engineering and computer science | Engineering, Computer science | Aerospace engineering, Biochemical engineering, Electrical engineering, Chemical engineering, Process engineering, Geotechnical engineering, Manufacturing engineering, Computer vision, Data mining, Computational science, Information retrieval, Computer security, Knowledge management, Civil engineering, Forensic engineering, Library science, Speech recognition, Operations research, Marine engineering, Reliability engineering, Mining engineering, Simulation, Telecommunications, Operating system, World Wide Web, Parallel computing, Systems engineering, Waste management, Transport engineering, Control engineering, Architectural engineering, Mechanical engineering, Construction engineering, Automotive engineering, Pattern recognition, Engineering physics, Process management, Machine learning, Computer engineering, Programming language, Human-computer interaction, Computer network, Engineering ethics, Petroleum engineering, Aeronautics, Structural engineering, Theoretical computer science, Nuclear engineering, Computer architecture, Computer graphics (images), Pulp and paper industry, Database, Internet privacy, Natural language processing, Data science, Real-time computing, Distributed computing, Algorithm, Embedded system, Artificial intelligence, Engineering management, Agricultural engineering, Industrial engineering, Electronic engineering, Multimedia, Computer hardware, Software engineering, Engineering drawing. |
| Life science | Environmental science, Medicine, Biology | Environmental planning, Molecular biology, Oncology, Virology, Bioinformatics, Environmental health, Medical emergency, Urology, Pathology, Biological system, Immunology, Cancer research, Botany, Physical medicine and rehabilitation, Dermatology, Biochemistry, Pharmacology, Animal science, Soil science, Andrology, Agricultural science, Gastroenterology, Ophthalmology, Paleontology, Biotechnology, Food science, |
| | | Toxicology, Optometry, Orthodontics, Genetics, Risk analysis (engineering), Gerontology, Internal medicine, Cardiology, Neuroscience, Family medicine, Veterinary medicine, Microbiology, Medical education, Medical physics, Physiology, Surgery, Dentistry, Agronomy, Zoology, Biomedical engineering, Cell biology, Ecology, Psychiatry, Obstetrics, Astrobiology, Horticulture, Environmental protection, Traditional medicine, Gynecology, Clinical psychology, Computational biology, Evolutionary biology, Anatomy, Intensive care medicine, Audiology, Biophysics, General surgery, Radiology, Pediatrics, Water resource management, Physical therapy, Agroforestry, Nursing, Environmental engineering, Anesthesia, Environmental resource management, Fishery, Nuclear medicine, Endocrinology, Emergency medicine. |
| Mathematics and physical science | Physics, Geography, Chemistry, Materials science, Geology, Mathematics, Statistics | Earth science, Geochemistry, Hydrology, Environmental chemistry, Particle physics, Applied mathematics, Combinatorics, Mathematical analysis, Analytical chemistry, Condensed matter physics, Photochemistry, Oceanography, Cartography, Algebra, Pure mathematics, Nuclear chemistry, Quantum mechanics, Composite material, Mechanics, Astronomy, Crystallography, Inorganic chemistry, Polymer chemistry, Nanotechnology, Forestry, Physical geography, Combinatorial chemistry, Discrete mathematics, Mathematics education, Atomic physics, Petrology, Arithmetic, Theoretical physics, Geometry, Quantum electrodynamics, Statistical physics, Computational chemistry, Archaeology, Economic geography, Nuclear magnetic resonance, Control theory, Polymer science, Seismology, Calculus, Mathematical physics, Stereochemistry, Classical mechanics, Astrophysics, Medicinal chemistry, Metallurgy, Geodesy, Acoustics, Remote sensing, Mathematical optimization, Topology, Meteorology, Statistics, Optics, Radiochemistry, Molecular physics, Nuclear physics, Computational physics, Chemical physics, Geophysics, Optoelectronics, Climatology, Geomorphology, Physical chemistry, Organic chemistry, Chromatography, Thermodynamics, Mineralogy, Ceramic materials, Atmospheric sciences, Biostatistics. |
| Social sciences and others | Art, Sociology, Economics, Political science, Philosophy, History, Psychology, Business | Art history, Commerce, Environmental ethics, Environmental economics, Social psychology, Aesthetics, International trade, Finance, Economic system, Gender studies, Psychoanalysis, International economics, Econometrics, Welfare economics, Financial economics, Ethnology, Social science, Socioeconomics, Applied psychology, Political economy, Management science, Economy, Visual arts, Marketing, Keynesian economics, Genealogy, Accounting, Literature, Regional science, Industrial organization, Demographic economics, Agricultural economics, Business administration, Management, Operations management, |

| | | Classics, Mathematical economics, Anthropology, Media studies, Criminology, Actuarial science, Linguistics, Development economics, Economic history, Pedagogy, Public administration, Public economics, Market economy, Public relations, Positive economics, Demography, Humanities, Natural resource economics, Psychotherapist, Religious studies, Theology, Economic policy, Advertising, Ancient history, Monetary economics, Economic growth, Financial system, Neoclassical economics, Law and economics, Law, Communication, Epistemology, Labor economics, Cognitive psychology, Classical economics, Microeconomics, Cognitive science, Developmental psychology, Macroeconomics. |
|---|---|---|

We leveraged Google Maps API to parse all 27,077 institution names in Microsoft Academic Graph, and retrieved their country labels. Therefore, we could label every Chinese scientist's working country in any publishing year. Specifically, we focused on Chinese scientists leaving the US, i.e., those who were trained in the US (first paper affiliated in the US) and who subsequently moved from the US to China (i.e., stopped using US affiliations and started to use Chinese affiliations). For each such scientist, we counted the year range of all his/her papers affiliated in the US and affiliated in China, and annotated his/her leaving year as the year of his/her first subsequent paper after his/her most recent usage of a US affiliation. This was more accurate than simply using his/her last year with a US affiliation, which might produce false positives that counted current US-based Chinese scientists. We further identified two groups of interest among US-based Chinese scientists: "junior" scientists—those who had published their first papers in the US, started publishing with Chinese affiliations within 5 years thereafter, and finally left the US within 7 years thereafter; and "experienced" scientists—those who had published over 25 papers in their whole career and outperformed 97% of scientists. Table S4 reports the yearly total number of US-based Chinese scientists who dropped US affiliation in each year since 2000. In Figures S1 to S3, we present the normalized trends for the groups as a whole and for the junior and experienced scientists.

Table S4: Yearly number of US-based Chinese scientists who dropped US affiliations for China affiliations

|  | Engineering and computer science | Mathematics and physical science | Life science | Social sciences |
|---|---|---|---|---|
| 2000 | 3 | 18 | 6 | 0 |
| 2001 | 3 | 19 | 12 | 0 |
| 2002 | 6 | 19 | 10 | 1 |
| 2003 | 15 | 25 | 24 | 2 |
| 2004 | 18 | 44 | 37 | 2 |
| 2005 | 22 | 49 | 36 | 5 |
| 2006 | 30 | 74 | 51 | 6 |
| 2007 | 31 | 66 | 71 | 5 |
| 2008 | 36 | 111 | 81 | 11 |
| 2009 | 57 | 117 | 110 | 13 |
| 2010 | 77 | 149 | 131 | 13 |
| 2011 | 85 | 185 | 141 | 18 |
| 2012 | 85 | 193 | 192 | 20 |
| 2013 | 97 | 211 | 253 | 28 |
| 2014 | 138 | 239 | 280 | 34 |
| 2015 | 110 | 279 | 294 | 32 |
| 2016 | 168 | 319 | 348 | 40 |
| 2017 | 175 | 341 | 348 | 50 |
| 2018 | 196 | 416 | 393 | 57 |
| 2019 | 202 | 468 | 430 | 73 |
| 2020 | 244 | 495 | 423 | 53 |
| 2021 | 298 | 639 | 478 | 75 |

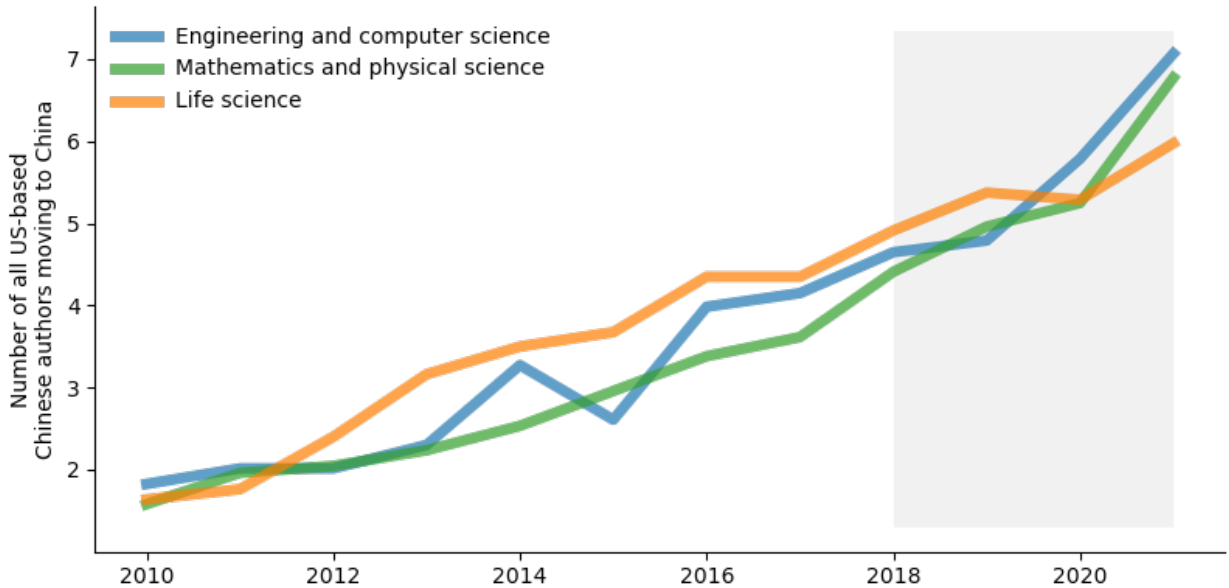

Figure S1: Trends in Chinese scientists migrating from the US to China. Number of all Chinese scientists leaving the US in each year from 2010 to 2021, normalized as ratios to the 2005–2010 level in each discipline.

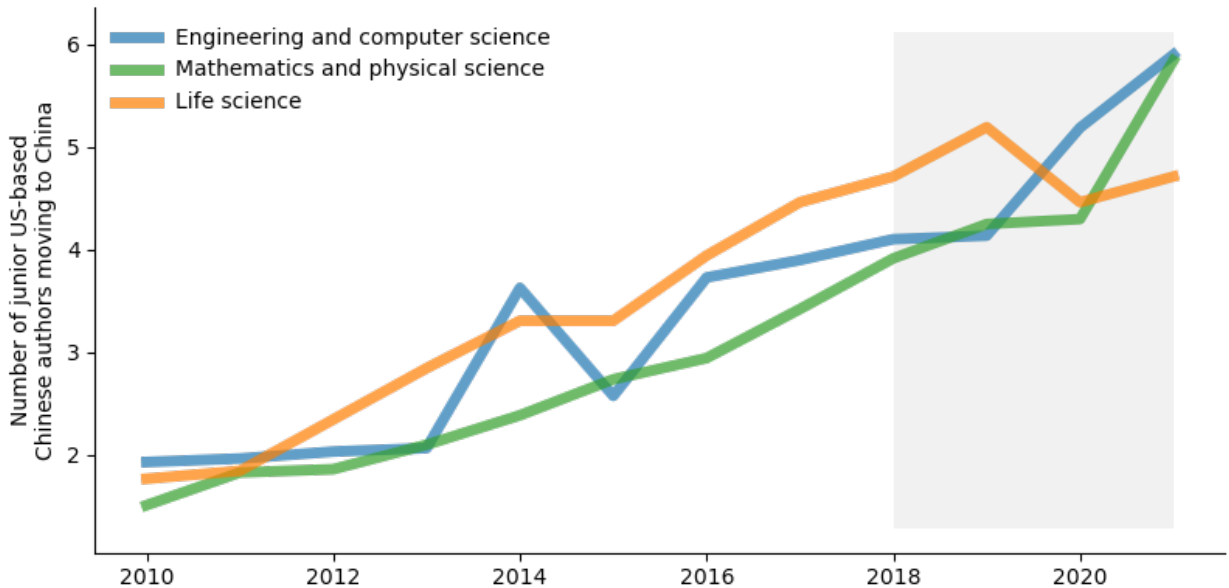

Figure S2: Trends in junior Chinese scientists migrating from the US to China. Number of junior Chinese scientists leaving the US in each year from 2010 to 2021, normalized as ratios to the 2005–2010 level in each discipline.

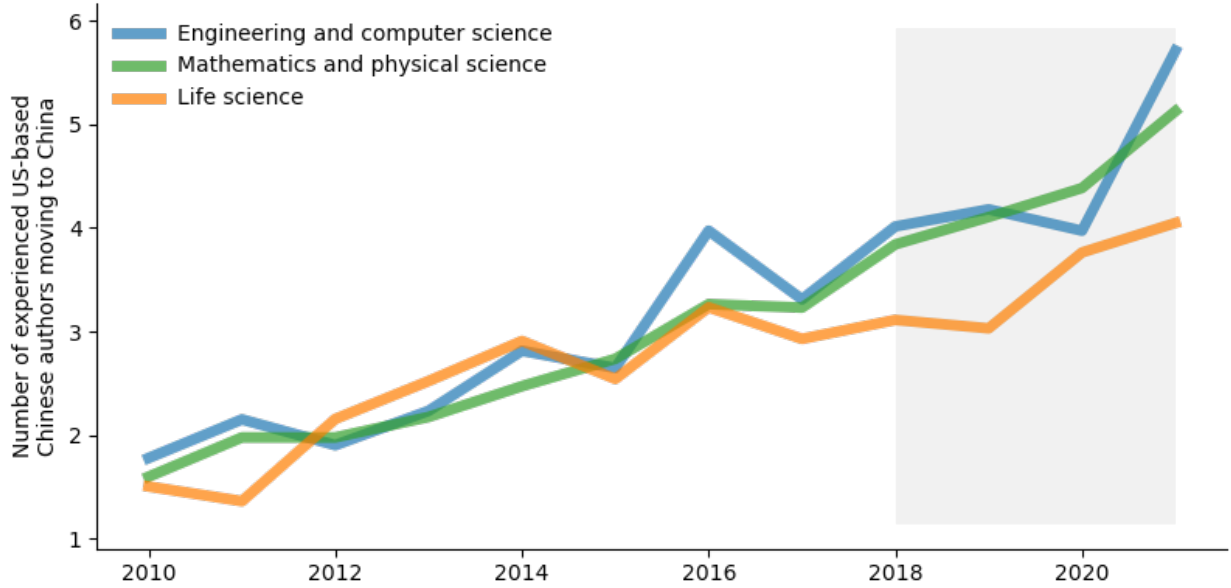

Figure S3: Trends in experienced Chinese scientists migrating from the US to China. Number of experienced Chinese scientists leaving the US in each year from 2010 to 2021, normalized as ratios to the 2005–2010 level in each discipline.

**Supplementary Materials 4: Asian American Academic Climate Survey**

The Asian American Scholar Forum (AASF) (aasforum.org) conducted an online survey of Asian American faculty in the US between December 2021 and March 2022. The stated objective of the survey was to understand challenges and experiences of Asian American scholars in their research and educational environments, including their perceptions of academic climate, academic activities, and mental and physical well-being. We designed the survey questionnaire and helped field the survey. To protect the confidentiality of the respondents, the survey began with a consent form and a promise that their responses would be collected and analyzed anonymously. In collaboration with various professional associations, the AASF sent the survey to intended respondents nationwide. Specifically, AASF asked all of its 55 members to forward the invitation message with the link to the survey to Chinese-American faculty members; AASF emailed the presidents of the following 11 Chinese-American professional associations that co-sponsored the AASF webinar series, asking them to forward the survey to their members (see below for the list).

1. Association of Chinese Scholars in Computing
2. Chinese-American Chemistry & Chemical Biology Professors Association
3. Chinese-American Oceanic and Atmospheric Association
4. Chinese Biological Investigator Society (CBIS)
5. International Chinese Statistical Association
6. North America Federation of Tsinghua Alumni Associations
7. Peking University Alumni Association of New England
8. Peking University Alumni Association of Washington DC
9. The Society of Chinese Bioscientists in America (SCBA)
10. Tsinghua Alumni Academia Club of North America
11. US Chinese Scholar Association of Combustion Institute

We obtained valid responses from 1,394 respondents. All participants signed the consent forms. For the analyses reported in this paper, we excluded 37 observations who self-identified as non-Chinese Asian American and 18 observations of missing Asian ethnicity. We further excluded 5 observations from graduate students and 30 observations for whom the current position was either missing or in industry. The aforementioned case exclusion criteria left us with a total of 1,304 observations of Chinese-American faculty. We then excluded 75 cases containing any missing values in the covariates for the logistic regression analysis presented in Table S5. Therefore, the main analytic sample size for predicting scholars' intention of relocating outside the US is 1,229, and the analytic sample size for predicting scholars' intention of avoiding federal grant applications is 934 (further restricted to those who had even been awarded grants from US government agencies). In Table S5, we provide the main descriptive statistics from the survey.

Methodologically, two sources of potential bias could be present in the AASF survey. The first is called "sample selection bias": potential respondents were more likely to participate in the AASF survey if they already perceived themselves to have been impacted by the China Initiative. The second is called "social desirability bias": respondents knew the objective of the AASF survey and may have supplied information consistent with the objective. Note that both sources of bias are in the direction of exaggeration of the negative impact of the China Initiative. Therefore, the results reported in this article should be interpreted with caution.

**TABLE S5. Descriptive Summary for the Main Analytical Sample of AASF Survey Data (n=1299)**

|  | Percentages |
|---|---|
| **Intention of Relocating Abroad (Either Asian or non-Asian Countries)** | 61% |
| Intention to relocate to Asian countries | 47% |
| Intention to relocate to non-Asian countries | 46% |
| **Intention of Avoiding Federal Grants[1]** | 45% |
| **Have Been Awarded a US Federal Grant** | 77% |
| **Intention of Contributing to the US Leadership in Science and Technology** | 89% |
| **Perceptions of Current Academic Climate:** | |
| Feel unwelcome as an academic researcher in the US | 35% |
| Do not feel safe as an academic researcher in the US | 72% |
| Fearful of conducting research in the US | 42% |
| Worried about collaborations with China | 65% |
| It is more difficult to recruit top international students now | 86% |
| Received disclosure inquiries from my institution in the last two years | 42% |
| **Sense of Belonging to Local Institution and Professional Community:** | |
| Feel that I belong | 55% |
| Neutral | 28% |
| Feel that I don't belong | 17% |
| **How Often Have You Been Bullied under Professional Settings Last Year?** | |
| Never | 25% |
| Rarely/Sometimes | 59% |
| Often/Most of the time | 10% |
| Not Sure | 6% |
| **How Often Have You Been Insulted by Others under Non-professional Settings Last Year?** | |
| Never | 13% |
| Rarely/Sometimes | 72% |
| Often/Most of the time | 11% |
| Not Sure | 4% |
| **Current Position:** | |
| Assistant Professor | 24% |
| Associate Professor | 23% |
| Full Professor | 48% |
| Non-tenure-track academic | 5% |
| **Male** | 74% |
| **Field of Study:** | |
| Mathematics and physical science | 29% |
| Life Science | 30% |
| Engineering and computer Science | 35% |
| Social Sciences and others | 6% |
| **Region of Institution:** | |
| West | 19% |
| Midwest | 24% |
| Northeast | 21% |
| South | 37% |
| **Type of Institution:** | |
| Public | 70% |
| Private | 30% |

Note: Based on the larger analytic sample focusing on intentions of relocating abroad.

   1: Among those ever-awardees of grants from US government agencies, 45% indicated intentions to avoid federal grants.

## Supplementary Materials 5: Explaining Stated Intentions

**TABLE S6. Logistic Regression Models Predicting Scholars' Intentions of Avoiding Applying for Federal Grants and of Relocating Abroad**

| | Scholar Intentions | | | |
| --- | --- | --- | --- | --- |
| | Avoiding Federal Grants [1] | | Relocating Abroad | |
| | Model 1A | Model 2A | Model 1B | Model 2B |
| **Perceptions of Current Academic Climate:** | | | | |
| Feel unwelcome as an academic researcher in the US | | 0.465* | | 0.505** |
| | | (0.189) | | (0.173) |
| Do not feel safe as an academic researcher in the US | | 0.807*** | | 0.727*** |
| | | (0.219) | | (0.159) |
| Fearful of conducting research in the US | | 1.389*** | | 0.523** |
| | | (0.187) | | (0.166) |
| Worried about collaborations with China | | 0.794*** | | 0.529*** |
| | | (0.192) | | (0.148) |
| It is more difficult to recruit top international students now | | 0.493+ | | 0.535** |
| | | (0.274) | | (0.193) |
| Received disclosure inquiries from my institution in the last two years | | -0.206 | | 0.029 |
| | | (0.166) | | (0.140) |
| **Sense of Belonging to Local Institution and Professional Community (ref. Feel that I belong):** | | | | |
| Neutral | | -0.154 | | 0.392* |
| | | (0.196) | | (0.161) |
| Feel that I don't belong | | 0.234 | | 0.412+ |
| | | (0.269) | | (0.223) |
| **How Often Have You Been Bullied under Professional Settings Last Year? (ref. Never)** | | | | |
| Rarely/Sometimes | | 0.098 | | 0.091 |

| | | | |
|---|---|---|---|
| | | (0.230) | (0.185) |
| Often/Most of the time | | 0.864* | 0.244 |
| | | (0.424) | (0.361) |
| Not sure | | 0.550 | 0.208 |
| | | (0.443) | (0.327) |
| **How Often Have You Been Insulted by Others under Non-professional Settings Last Year? (ref. Never):** | | | |
| Rarely/Sometimes | | 0.165 | 0.793*** |
| | | (0.300) | (0.230) |
| Often/Most of the time | | 0.151 | 1.134** |
| | | (0.424) | (0.361) |
| Not sure | | -0.477 | 0.614 |
| | | (0.567) | (0.428) |
| **Current Position (ref. Full Professor):** | | | |
| Assistant Professor | -0.779*** | -0.642** | 0.517*** | 0.925*** |
| | (0.190) | (0.225) | (0.157) | (0.184) |
| Associate Professor | -0.223 | -0.045 | 0.368* | 0.593*** |
| | (0.168) | (0.201) | (0.153) | (0.175) |
| Non-tenure-track academic | -0.485 | 0.210 | -0.047 | 0.289 |
| | (0.398) | (0.460) | (0.271) | (0.311) |
| **Male (ref. Female)** | 0.325+ | 0.089 | 0.236+ | 0.072 |
| | (0.167) | (0.200) | (0.139) | (0.159) |
| **Field of Study (ref. Mathematics and physical science):** | | | |
| Life science | -0.330+ | -0.792*** | 0.068 | -0.191 |
| | (0.180) | (0.218) | (0.157) | (0.179) |
| Engineering and computer science | 0.493** | 0.128 | -0.052 | -0.384* |
| | (0.170) | (0.202) | (0.149) | (0.170) |
| Social Sciences/Others | 0.455 | -0.090 | 0.387 | 0.317 |
| | (0.512) | (0.627) | (0.280) | (0.314) |

**Region of Institution (ref. West)**

|  |  |  |  |  |
|---|---|---|---|---|
| Midwest | 0.215 | 0.338 | 0.027 | 0.012 |
|  | (0.210) | (0.247) | (0.184) | (0.205) |
| Northeast | -0.043 | 0.077 | -0.164 | -0.085 |
|  | (0.224) | (0.262) | (0.194) | (0.216) |
| South | -0.022 | 0.178 | 0.076 | 0.102 |
|  | (0.192) | (0.229) | (0.170) | (0.190) |
|  |  |  |  |  |
| **Public Institution (ref. Private Institution)** | 0.429** | 0.474* | 0.138 | 0.079 |
|  | (0.161) | (0.191) | (0.139) | (0.156) |
|  |  |  |  |  |
| **Have Been Awarded a US Federal Grant (Ref. Never)** |  |  | 0.433** | 0.494** |
|  |  |  | (0.154) | (0.178) |
|  |  |  |  |  |
| Constant | -0.681** | -3.084*** | -0.362 | -2.871*** |
|  | (0.259) | (0.474) | (0.268) | (0.394) |
|  |  |  |  |  |
| Observations | 936 | 934 | 1,234 | 1,229 |
| Pseudo R2 | 0.0503 | 0.259 | 0.0168 | 0.167 |

Notes: 1. the analytic sample for "avoiding federal grants" is restricted to those ever-awardees (past or current) of grants from US government agencies.
Reporting the coefficients from logistic regression models; standard errors in parentheses; *** $p<0.001$, ** $p<0.01$, * $p<0.05$, + $p<0.10$.

# Supplementary Materials 6: Predicting Fears

**TABLE S7. Logistic Regression Models Predicting Scholars' Fears**

| | Indicators of Fear | | |
| --- | --- | --- | --- |
| | Do Not Feel Safe | Feel Unwelcome | Fearful of Conducting Research |
| | Model 1 | Model 2 | Model 3 |
| **Current Position (ref. Full Professor):** | | | |
| Assistant Professor | -0.233 | -0.198 | -0.525*** |
| | (0.166) | (0.157) | (0.157) |
| Associate Professor | -0.180 | -0.113 | -0.136 |
| | (0.164) | (0.153) | (0.150) |
| Non-tenure Track Academic | -0.455 | -0.234 | -0.733* |
| | (0.286) | (0.293) | (0.306) |
| **Field of Study (ref. Mathematics and physical science):** | | | |
| Life science | 0.144 | 0.307+ | 0.603*** |
| | (0.166) | (0.163) | (0.160) |
| Engineering and computer science | 0.522** | 0.414** | 0.743*** |
| | (0.165) | (0.154) | (0.153) |
| Social Sciences/Others | -0.313 | 0.344 | 0.385 |
| | (0.275) | (0.281) | (0.289) |
| **Male (ref. Female)** | 0.218 | 0.219 | 0.389** |
| | (0.147) | (0.145) | (0.143) |
| **Region of Institution (ref. West)** | | | |
| Midwest | -0.187 | 0.020 | -0.067 |
| | (0.204) | (0.185) | (0.184) |
| Northeast | -0.254 | -0.083 | -0.044 |
| | (0.214) | (0.201) | (0.197) |
| South | -0.200 | -0.033 | -0.116 |
| | (0.188) | (0.171) | (0.170) |
| **Public Institution (ref. Private Institution)** | 0.143 | 0.329* | 0.190 |
| | (0.150) | (0.145) | (0.141) |
| **Have Been Awarded a US Federal Grant (Ref. Never)** | -0.019 | -0.000 | 0.346* |
| | (0.166) | (0.159) | (0.161) |
| Constant | 0.808** | -1.156*** | -1.240*** |
| | (0.289) | (0.279) | (0.278) |
| | | | |
| Observations | 1,234 | 1,234 | 1,234 |
| Pseudo R2 | 0.0195 | 0.0133 | 0.0431 |

Reporting the coefficients from logistic regression models; standard errors in parentheses.

*** p<0.001, ** p<0.01, * p<0.05, + p<0.10.

## Supplementary Materials 7: Explaining Fears

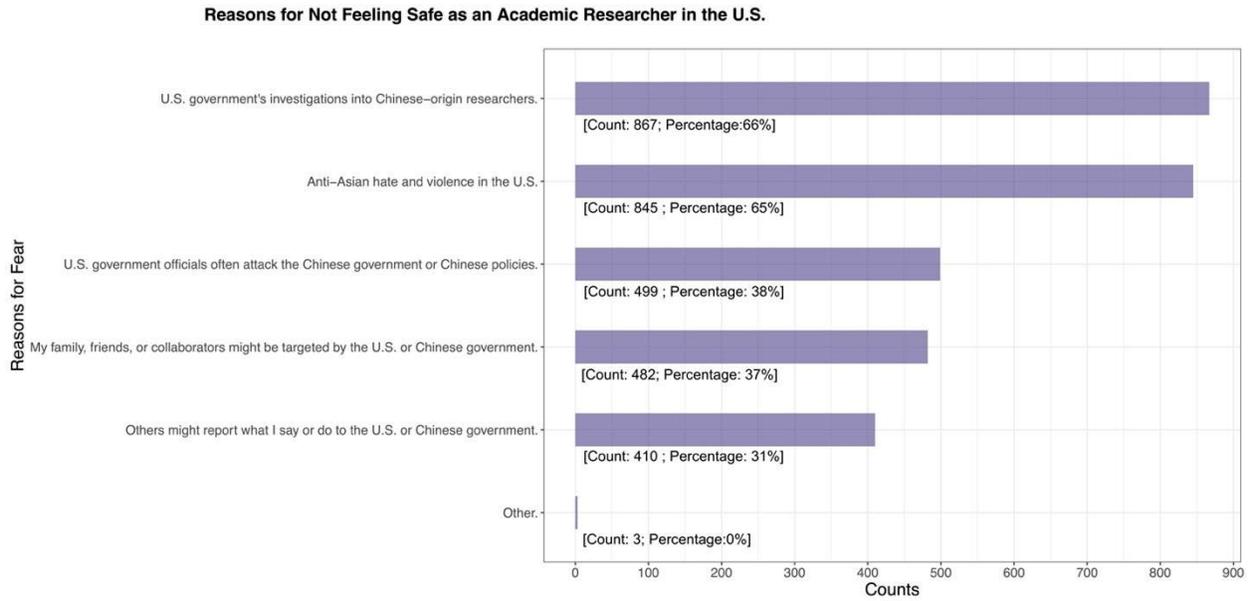

Figure S4: Reasons for not feeling safe as an academic researcher in the US

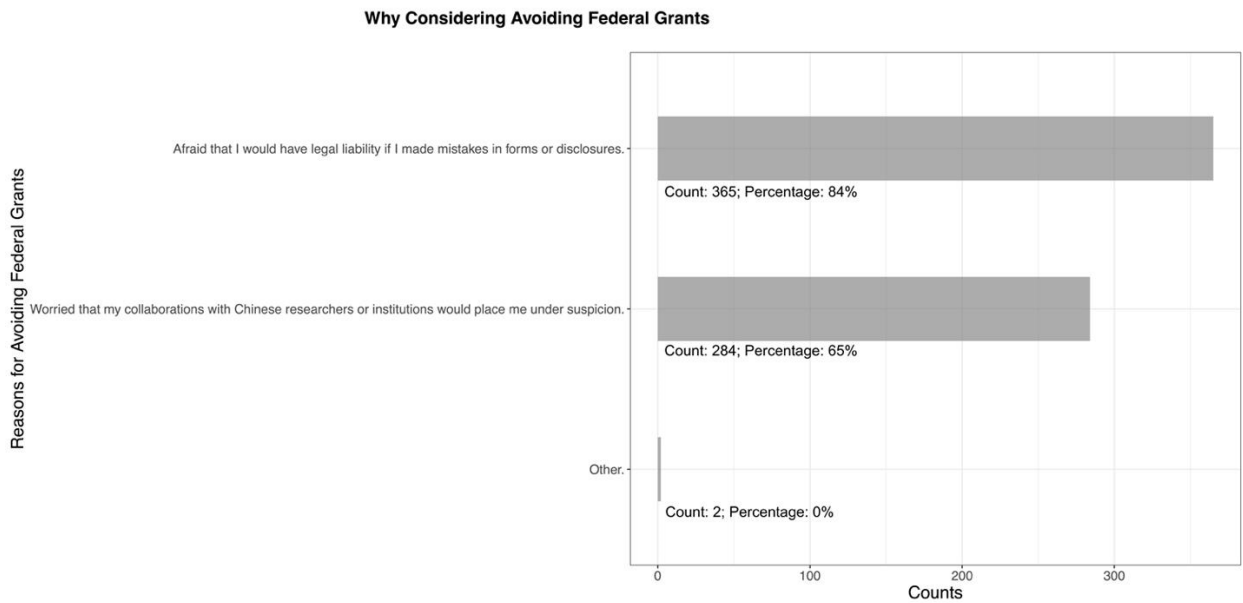

Figure S5: Reasons for considering avoiding applying for federal grants (N=436)

**Supplementary Materials 8:  Comparison of the AASF Survey to Two Other Surveys**

Two additional surveys on the same topic were conducted: the University of Arizona survey and the University of Michigan survey. For simplicity, we will refer to the first survey as the UA survey and the second survey as the UM survey. In Table S8, we compare the sampling methods of the three surveys in detail, using sources in (*6,7*). By survey standards, all of the three surveys are considered "convenience" samples. That is, they are not probability-based (which is the most desirable) samples because there is no national sampling frame from which a sample could be drawn. In addition, we do not know the response rate of the AASF survey. Because the AASF survey is a convenience sample with an unknown response rate, we acknowledge that the results can be subject to sampling and response biases.

Table S8: Comparison of Methodology across the Three Surveys

| | AASF Survey | University of Michigan (UM) Survey | University of Arizona (UA) Survey |
|---|---|---|---|
| Survey Distribution Methods | (1) AASF asked all its 55 members to forward the invitation message with the link to the survey to Chinese-American faculty members in their networks; (2) AASF emailed to the presidents of the 11 Chinese-American professional associations that co-sponsored the AASF webinar series, asking them to forward the survey to their members (see below for the list). | "Invitations were sent to 927 members of Asian/Chinese faculty associations at five universities: University of Michigan, Michigan State, Iowa State, Columbia, and Notre Dame" (according to the slides shared by the University of Michigan survey team). | (1) "The University of Arizona and the Committee of 100 administered a national survey between May and July 2021 among scientists in top US universities, including faculty, post-doctoral fellows (postdocs), and graduate students. … The survey was sent to: a) all Chinese name scientists; and b) a random sample of non-Chinese name scientists across 83 US universities. … (2) In order to purposely oversample Chinese scientists for comparison, we sent the survey invitation through email to the entire Chinese name group, and an equivalent number of randomly selected scientists from the non-Chinese name group." (*7*, p.29) |
| Sample Size | 1,394 valid responses in total, including 1,304 Chinese-American faculty members (unknown response rate). | 295 full responses (32% response rate). | 1,060 responses from scientists with Chinese surnames and 889 responses from scientists with non-Chinese surnames.  Total sample size is 1,949 (6.8% overall response rate). |
| Survey-Fielding Dates | December 2021–March 2022 | July–August 2021 | May–July 2021 |
| Sample Composition | Chinese faculty members at US institutions nationwide (excluding students) | Asian/Chinese faculty members at five institutions (excluding students and postdocs) | Chinese and non-Chinese faculty members, including postdocs and graduate students. |

Note: Sources for the UM survey and the UA survey and are in (*6, 7*). List of associations that forwarded the AASF survey invitation: Association of Chinese Scholars in Computing, Chinese-American Chemistry & Chemical Biology Professors Association, Chinese-American Oceanic and Atmospheric Association, Chinese Biological Investigator Society (CBIS), International Chinese Statistical Association, North America Federation of Tsinghua Alumni Associations, Peking University Alumni Association of New England, Peking University Alumni Association of Washington DC, The Society of Chinese Bioscientists in America (SCBA), Tsinghua Alumni Academia Club of North America, US Chinese Scholar Association of Combustion Institute.

We further compare the main findings from the three surveys, summarized in Table S9. Because the AASF survey is primarily a survey of Chinese-origin academic scientists, we compare the results to those of "Chinese" scientists in the UA survey. Table S9 shows that all major findings, when they are comparable, are remarkably consistent across the three surveys. With different wordings for the question on feelings of safety, for example, 51% of the respondents in the AASF survey feel unsafe, 59% of the respondents in the UM survey do not feel safe, and 50.7% of the Chinese respondents in the UA survey feel fear/anxiety of being surveilled by the US government. The three surveys each collected information on respondents' feelings toward applying for federal grants. In the AASF full analytical sample, 34% have considered avoiding applications for federal grants due to the current political climate in the US; in the UM sample, 28% have considered avoiding applying for federal grants; in the UA sample, 38.4% report having experienced more difficulty in obtaining research funding in the US as a result of their race/nationality/country of origin. Responses concerning intentions to leave the US are also consistent across the surveys: In the AASF survey, 46% intend to relocate to Asia, and 47% to non-Asian countries; in the UM survey, 32.2% have thought about moving to Asia, and 26.2% to Canada, Europe, Australia, or New Zealand; in the UA survey, 42.1%. report that FBI investigations and the China Initiative have affected their plans to stay in the US. Similar consistency is found for other survey items of interest when they are comparable across the surveys.

Table S9: Comparison of Findings from the Three Surveys

| | AASF Survey | University of Michigan (UM) Survey | University of Arizona (UA) Survey (Chinese only) |
|---|---|---|---|
| Question: Do you feel safe… | I currently feel safe as an academic researcher in the US.<br><br>  Feel unsafe: 51%; unsure: 21%. | Do you feel safe as Chinese-origin academic researchers in the US?<br><br>  Do not feel safe: 59%; not sure: 12%. | Scientists who feel fear/anxiety of being surveilled by US gov't<br><br>  50.7%. |
| Question: Reasons for not feeling safe… | I do not feel safe because…<br><br>  Because of the US gov't investigations into Chinese-origin researchers: 66%.<br><br>  Because of anti-Asian violence in the US: 65%.<br><br>  Because US gov't officials often attack the Chinese gov't or Chinese policies: 38%.<br><br>  Because my family, friends, or collaborators might be targeted by <u>the U.S. or Chinese gov't</u>: 37%.<br><br>  Because <u>others</u> might report what I say or do in the US to Chinese gov't: 31%. | I do not feel safe because…<br><br>  Because of the US gov't investigations into Chinese-origin researchers: 56%.<br><br>  Because of anti-Asian violence in the US: 55.9%.<br><br>  Because US gov't officials often attack the Chinese gov't or Chinese policies: 29.4%<br><br>  Because <u>Chinese gov't</u> could target my family/friends/collaborators to retaliate: 10.7%.<br><br>  Because <u>other Chinese</u> might report what I say or do in the US to Chinese gov't: 8%. | |
| Question: Research grants (broadly defined) | Have you considered <u>avoiding applications for federal grants</u> due to the current political climate in the US?<br><br>Yes, I have:  34% of the full analytic sample (N=1234); 45% of ever-awardees of federal grants (n=936). | Have you considered <u>avoiding federal grants</u>?<br><br><br>Yes, I have: 28% of the full analytic sample (N=295). | Scientists who experience more difficulty in obtaining <u>research funding in the US</u> as a result of their race/nationality/country of origin<br><br>  38.4%. |

| Question: Intention to leave the US | Intention of relocating abroad (either Asian or non-Asian Countries):<br><br>61% overall. To Asia: 46%; to non-Asian countries: 47%. | Given current political environment in the US, thought about moving…<br><br>To Asia: 32.2%; to Canada, Europe, Australia or New Zealand: 26.2%. | Scientists who report that FBI investigations and/or the China Initiative affected their plans to stay in the US 42.1%. |
|---|---|---|---|
| Question: whether my university encouraged collaboration with China | Before 2018, did you feel that the University encouraged collaboration in China?<br><br>80% of the 922 non-missing responses (56% of the full analytic sample). | Before 2018, did you feel that the University encouraged collaboration in China?<br><br>77% of the 139 non-missing responses (i.e., 36% of the full survey sample). | |
| Question: whether my university still encourages collaboration with China now | Do you feel that this university currently encourages collaborations in China?<br><br>3.4% of the 916 non-missing responses (2.4% of the full analytic sample). | Do you feel that this university currently encourages collaborations in China?<br><br>9% of the 168 non-missing responses (5% of the full survey sample). | |

Sources:  Based on (1) our calculations from the AASF analytic sample; (2) PowerPoint slides shared by the UM Survey research team via email (*6*); and (3) public report of the UA Survey posted on Committee of 100 website (*7*).

Note:  In this table, we underlined comparable yet not identical questions asked across the three surveys.

**Supplementary Materials 9: Evaluation of the AASF Survey Using ACS Data**

Because the AASF survey is a convenience sample, it may not be representative of its underlying population. To evaluate the representativeness of the AASF sample, we compare a few key sociodemographic characteristics of the AASF sample to the American Community Survey (ACS), the "gold standard" government survey conducted by the US Census Bureau (https://www.census.gov/programs-surveys/acs/). Unfortunately, there are only a limited number of variables that are available in both the AASF survey and the ACS survey (pooled annual files 2015–2019). We present the results of the evaluation in Table S10. Note that the sample size of the ACS survey is small due to the sample restriction. There are some small discrepancies. For example, we observe a higher proportion of respondents in engineering and computer science, and a lower proportion in life science, in the AASF survey than in the ACS. One possibility is that a high proportion of Chinese-origin life scientists are employed in non-tenure-track positions and thus were non-eligible for the AASF survey. Engineers and computer scientists are likely to be employed in tenure-track positions and are eager to participate in the AASF survey because they are impacted by the China Initiative. Further, the AASF sample is much older than the ACS sample. Compared to younger researchers, senior researchers are more likely to be approached by professional organizations to participate in the AASF survey, and they are more motivated to participate in the survey because they are more likely to be impacted by the China Initiative. Aside from these two discrepancies, the demographic representativeness of the AASF survey is overall adequate.

Table S10: Comparison of Demographic Characteristics between the AASF Survey and the American Community Survey (ACS) 2015–2019

| | Main Analytical Sample of AASF Survey Data (n=1299, from TABLE S5) | ACS 2015–19, Pooled Annual Samples (n=662) |
|---|---|---|
| **Male** | 74% | 61% |
| **Field of Study:** | | |
| Formal/Physical Science and Statistics | 29% | 29% |
| Life Science | 30% | 46% |
| Engineering and Computer Science | 35% | 18% |
| Social Sciences/Others | 6% | 7% |
| **Region of Institution:** | | |
| West | 19% | 24% |
| Midwest | 24% | 18% |
| Northeast | 21% | 24% |
| South | 37% | 33% |
| **Age Category:** | | |
| 18–40 | 30% | 63% |
| 41–50 | 33% | 20% |
| 51–60 | 28% | 12% |
| 61+ | 9% | 4% |

Notes: The pooled sample of American Community Survey (ACS 2015–19) is restricted to foreign-born respondents aged 18+, whose race is "Chinese," holding "doctoral degree," whose industry is "colleges and universities," and whose occupation is broadly defined as a "scientist." Unfortunately, we cannot further restrict the ACS sample to those who hold tenure-track positions versus non-tenure-track positions.